\documentstyle[epsf]{lamuphys}
\newcommand{\BEQ}{\begin{equation}}
\newcommand{\EEQ}{\end{equation}}
\newcommand{\BEA}{\begin{eqnarray}}
\newcommand{\EEA}{\end{eqnarray}}
\renewcommand{\H}{{\cal H}}
\newcommand{\tr}{{\rm tr}}
\newcommand{\N}{{\cal N}}
\renewcommand{\I}{{\cal I}}
\renewcommand{\D}{{\tilde D}}
\input{psfig}
 \begin{document}
\title{Complexity as the driving force for  glassy transitions}
 \author{Th.~M.~Nieuwenhuizen}
 \institute{
 Van der Waals-Zeeman Institute, Universiteit van Amsterdam\\ 
	 Valckenierstraat 65, 1018 XE Amsterdam, The Netherlands\\
e-mail: nieuwenh@phys.uva.nl}
 \date{\today}
\maketitle
\begin{abstract}
The glass transition is considered within 
two toys models, a mean field spin glass and a directed polymer
in a correlated random potential. 

In the spin glass model there occurs
a dynamical transition, where the 
the system condenses in a state of lower entropy.
The extensive entropy loss, called  complexity or information entropy, 
is calculated by analysis of the metastable (TAP) states.
This yields a well behaved thermodynamics of the dynamical transition.  
The multitude of glassy states also implies an extensive
difference between the internal energy fluctuations 
and the specific heat. 

In the directed polymer problem there occurs a thermodynamic
phase transition in non-extensive terms of the free energy.
At low temperature the polymer condenses in a set of highly
degenerate metastable states.
\end{abstract}
\section{Introduction}
The structural glass transition is said to occur at 
the temperature $T_g$ where the viscosity equals $10^{14}$ Poise.
The question why this transition occurs is often ``answered'' 
(more correctly: avoided) 
by saying that it is a  dynamical transition. Surely, there is a 
continuum of time scales
ranging from picoseconds to many years; at experimental 
time scales there is no equilibrium. Nevertheless,
since some 20 decades in time
are spanned, one would hope that equilibrium statistical mechanics
can be applied in some modified way.

Crudely speaking, the observation time will set a
scale. Processes with shorter timescales can be considered in equilibrium;
processes with longer timescales are essentially frozen, 
as if they were random.
To provide a (non-equilibrium) thermodynamic explanation of a model glassy
 transition will be the first subject of the present work.

Intuitively we expect that the resulting free energy is given by 
the logarithm of the partition sum, provided it has been restricted 
to those states 
that can be reached dynamically in the timespan considered. 
This non-equilibrium free energy will then differ from
the standard case, and need not be a thermodynamic potential that determines
the internal energy and entropy by its derivatives. 

Experimentally one often determines the entropy $S_{exp}$, and thus the
free energy $F_{exp}=U-TS_{exp}$, 
 from the specific heat data by integrating 
 $C/T$ from a reference temperature in the liquid phase down to $T$.  
As long as the cooling rate is finite there remains 
at zero temperature a residual entropy. In the limit of adiabatically
slow cooling it vanishes.

  Alternatively, a glass can be seen as a disordered solid.
 In this description the liquid undergoes a transition to a glass
 state with extensively smaller entropy.
These states are sometimes called ``states'',  ``metastable states'', 
``components'' or, in spin glass theory, ``TAP-states''.
 As the  free energy then becomes much larger, it is not so evident  from 
 thermodynamic considerations why the system can get captured in such a state
 with much and much smaller Gibbs-weight $\sim\exp(-$volume$)$.
 The explanation is that the condensed system then has lost part 
 of its entropy,  namely  the  entropy of selecting one out of the many 
 equivalent states. 
 This part, ${\cal I}$, is called the 
 {\it configurational entropy}, {\it complexity} 
 or {\it information entropy}~\cite{Jackle}~\cite{Palmer}.
 Its origin can be understood as follows.
 When the Gibbs free energy $F_{\bar a}$ of the relevant state $\bar a$ has a 
 large degeneracy ${\cal N}_{\bar a}\equiv \exp({\cal I}_{\bar a})$, the
 partition sum yields $Z=\sum_a\exp(-\beta F_a)\approx 
{\cal N}_{\bar a}\exp(-\beta F_{\bar a})$, so 
 $F=F_{\bar a}-T{\cal I}_{\bar a}$ is the full free energy of the system. 
 The entropy loss arises when the system chooses the state to condense into,
 since from then on only that single state is observed.~\cite{quantummeas}
 As the total entropy $S=S_{\bar a}+{\cal I}_{\bar a}$ is continuous, 
 so is the total free energy. For an adiabatic cooling experiment 
J\"ackle has  assumed that the weights $p_a$ of the states $a$ are
 fixed at the transition,\cite{Jackle} which implies that the free energy
 difference between the condensed phase and the liquid is positive and
 grows quadratically below $T_c$.
 This explains the well known discontinuity in quantities such as the 
 specific heat. 
 However, it may seem unsatisfactory that this higher free energy branch 
 describes the physical state.

We shall first investigate these questions for the dynamical transition 
of a mean field spin glass model, and then for the static transition of a 
directed polymer in a correlated potential.

\section{The $p$-spin glass}
 We first analyze these thermodynamic questions within a
relatively well understood spin glass model, the mean 
field $p$-spin interaction spin glass.
For a system with $N$ spins we consider the Hamiltonian 
\begin{equation}\label{Ham=}
{\cal H}=-\sum_{i_1<i_2<\cdots<i_p} J_{i_1 i_2\cdots i_p}
S_{i_1}S_{i_2}\cdots S_{i_p}
\end{equation}
with independent Gaussian random couplings, that have average zero
and variance $J^2p!/2N^{p-1}$.

  Kirkpatrick and Thirumalai\cite{KirkpT} pointed out 
  for the case of Ising spins that for $p=3$ there is a close analogy with
 models for the structural glass transition, and that its properties are
 quite insensitive of the value of $p$ as long as $p>2$. 

 The spherical limit of this model, where
 the spins are real valued but subject to the spherical
condition $\sum_i S_i^2=N$,  is very instructive. 
It  has received quite some attention recently.
The static problem was solved by Crisanti and Sommers.~\cite{CS} 
There occurs a static first order transition to a state with
 one step replica symmetry breaking (1RSB) at a temperature $T_g$.

 The dynamics of this spherical model was studied by 
Crisanti, Horner and Sommers 
 (CHS)~ \cite{CHS} and Cugliandolo and Kurchan~\cite{CK}. 
 Both groups find a sharp dynamical transition at a temperature $T_c>T_g$,
 which can be interpreted on a quasi-static level as a 1RSB transition.
This dynamical transition is sharp since in mean field 
the metastable states have infinite lifetime.
 For $T<T_c$ one of the fluctuation modes is massless (``marginal''),
 not unexpected for a glassy state.
 At $T_c^-$ there is a lower specific heat.

These dynamical approaches are the equivalent for the spin glass
of the mode coupling equations for the liquid-glass transitions.
~\cite{Gotze} At a critical
temperature $T_c>T_g$ a dynamic phase transition has been reported. 
The presence of a sharp  transition has been questioned, 
however.~\cite{Rudi}

 CHS integrate  $C/T$ to define the ``experimental''  entropy $S_{exp}$
 and the resulting free energy $F_{exp}=U-TS_{exp}$
 exceeds  the paramagnetic one  quadratically. 
 The interpretation of metastable  states (``TAP-states'') in this 
 system is discussed in ref. \cite{KPV} 
 The statistics of those states 
 was  considered by Crisanti and Sommers (CS).~\cite{CSTAP} 
 Assuming that  the result of long time dynamics follows through 
 being stuck in the metastable state of highest complexity, 
 they reproduced the ``experimental'' free energy obtained of {CHS}.
 This confirms J\"ackle's prediction of a quadratically 
higher free energy in the glassy state.

The long-time dynamics of 1RSB transitions fixes $q_0$ ($=0$ in zero field), 
$q_1$  and $x$, which 
are just the plateau values  and the breakpoint of a related Parisi order
parameter function, respectively. In $p$-spin models they
can simply be derived from a 1RSB replica  calculation provided 
one fixes $x$ by
a marginality criterion for fluctuations on the $q_1$ plateau.
~\cite{KirkpT}~\cite{CHS}~\cite{CK} 
The present author recently assumed that this is a very general phenomenon. 
~\cite{maxmin} This was motivated by the
expectation that a dynamical transition 
will automatically  get trapped in a state with diverging time scale, if 
present. 
In a Potts model this then predicts a dynamical transition with marginal
$q_0$ plateau and stable $q_1$ plateau.

As the replica free energy is minimized in this procedure, it  
lies  near $T_c^-$ {\it below} the paramagnetic value and  has 
a larger slope. Though this is exactly what one expects at a first order 
phase transition, it is a new result for dynamical glassy transitions.
It is the purpose of the present work to discuss the physical meaning
of the mentioned free energies.

\subsection{The replica free energy}

At zero field the 1RSB replica calculation involves the
plateau value $q_1$ and the breakpoint $x$. It yields the free 
energy~\cite{CS}
\begin{eqnarray}\label{bFCS}
\frac{ F_{repl}}{N}&=&-\frac{\beta J^2}{4}+\frac{\beta J^2}{4}
\xi q_1^p\\
&-&\frac{T}{2x}\log(1-\xi q_1)+\frac{T\xi}{2x}\log(1-q_1)
\nonumber\end{eqnarray}
where $\xi=1-x$. The first term describes 
the paramagnetic free energy.
Here and in the sequel, we omit the $T=\infty$ entropy. It is a constant,
only fixed after quantizing the spherical 
model,~\cite{Nqsg} that plays no role in the present discussion.
For the marginal solution $q_1$ is fixed by equating the lowest 
fluctuation eigenvalue to zero, which gives
\BEQ\label{marg}
\frac{1}{2}p(p-1)\beta^2 J^2q_1^{p-2}(1-q_1)^2=1.
\EEQ
 The condition
$\partial F/\partial q_1=0$ then yields $x=x(T)\equiv(p-2)(1-q_1)/q_1$.
This dynamical transition sets in at  temperature
$T_{c}=J \{p(p-2)^{p-2}/2(p-1)^{p-1}\}^{1/2}$ where $x$ comes below
unity. The same transition temperature follows from dynamics.

\subsection{Components}

A state, called a {\it component} by 
Palmer, ~\cite{Palmer} is labeled by
$a=1,2,\cdots,{\cal N}$, and has a local magnetization profile
$m_i^a=\langle S_i\rangle^a$. Its free energy 
$F_a$ is a thermodynamic potential that determines the internal energy
and the entropy by its derivatives. In the present model  
$F_a=F_{TAP}(m_i^a)$ is know explicitly. It is 
a minimum of the
``TAP'' free energy functional~\cite{Rieger}~\cite{KPV}~\cite{CSTAP}
\begin{eqnarray} \label{FTAP=}
&F&_{TAP}(m_i)=
-\sum_{i_1<\cdots<i_p}J_{i_1
\cdots i_p}m_{i_1}
\cdots m_{i_p} -H\sum_im_i
 \nonumber\\
&-&\frac{NT}{2}\log(1-q)
-\frac{N\beta J^2}{4}(1+(p-1)q^p-pq^{p-1})
\end{eqnarray}
where $q=(1/N)\sum_i m_i^2$ is the self-overlap. The state $a$ occurs with
weight $p_a$ that is set by the type of experiment one describes. 
(In practice these weights are usually unknown.)
Given the $p_a$'s one can define the 
``component averages'' such as $\overline F=\sum_a p_aF_a$,
 $\overline C=\sum_a p_aC_a$ and even the
 complexity ~\cite{Jackle}~\cite{Palmer}
 ${\cal I}=-\sum_a p_a \ln{p_a}$. 
For any observable, the component overage is the object
one obtains when measuring over repeated  runs 
and averaging over the outcomes. 
According to the Gibbs weight the probability of occurrence is
$p_a=\exp(-\beta F_a(T))/Z$, with $Z=\sum_a\exp(-\beta F_a)$.

The nice thing of the present model is that many questions can be 
answered directly. After setting $\partial F_{TAP}/\partial m_i=0$, 
we can use this equation to express $F_a$ in terms of $q_a$ alone. 
This gives the simple relation $F_a=Nf(q_a)$ where
\begin{eqnarray}\label{fiTAP=}
f(q)&=&\frac{\beta J^2}{4}[-1+(p-1)q^p-(p-2)q^{p-1}]\nonumber\\
&-&\frac{Tq}{p(1-q)}
-\frac{T}{2}\log(1-q) 
\end{eqnarray}
The resulting saddle point equation for $q_a$ coincides with the marginality
for $q_1$ given below eq. (2). 
Since $F_a$ only depends on the selfoverlap $q_a$, it 
is self-averaging. In the paramagnet one has $m_i=q=0$, so both eqs. 
(\ref{FTAP=}) and (\ref{fiTAP=}) reproduce the replica free energy
$F=-N\beta J^2/4$.  From the replica analysis we know that at
 $T_c^-$ the value of $q_a\approx q_1$ is $q_1=q_c\equiv (p-2)/(p-1)$.
The component free energy $F_a=Nf(q_c)$ exceeds the free energy of the
paramagnet by an extensive amount.
 As expected from experimental knowledge on glasses,
the internal energy is found to be continuous. 
At $T_c^-$ the free energy difference is solely due to the lower entropy, 
$S_a=-N\beta^2 J^2/4-{\cal I}_c$, where
\begin{equation} \label{I=}
{\cal I}_c=N\left(\frac{1}{2}\log(p-1)+\frac{2}{p}-1\right)
\end{equation}
is the value of complexity at the transition point.

This discussion 
supports the picture of the glass as a disordered solid, where
the entropy of the component the system condenses in, 
and thus the component average $\overline S$, 
is much smaller that the entropy of the paramagnet.
In real glasses this loss of entropy is due to the reduced phase 
space that arises by trapping of the atoms in a glassy configuration. 
In the quantized system ${\overline S}$ will vanish at $T=0$.~\cite{Nqsg}

\subsection{Value of the complexity}

 Kirkpatrick, Thirumalai and Wolynes ~\cite{KWol} 
 were the first to study the role of the complexity for 
Potts glasses in static situations in the temperature range $T_g<T<T_c$, 
see also ~\cite{KirkpT} for Ising spin glasses.
Statically ( that is to say, on timescales $\sim \exp(N)$) 
the system condenses into a state with higher free energy 
but with complexity such that the
total free energy is exactly equal to the paramagnetic free energy.
Here we will investigate the role of the complexity 
on timescales $\sim N^\gamma$,
 relevant for the dynamical transition at $T_c$.

The free energies discussed for this problem are plotted in Figure 1.

\begin{figure}[tbh]
\label{CSclfig}
\psfig{file=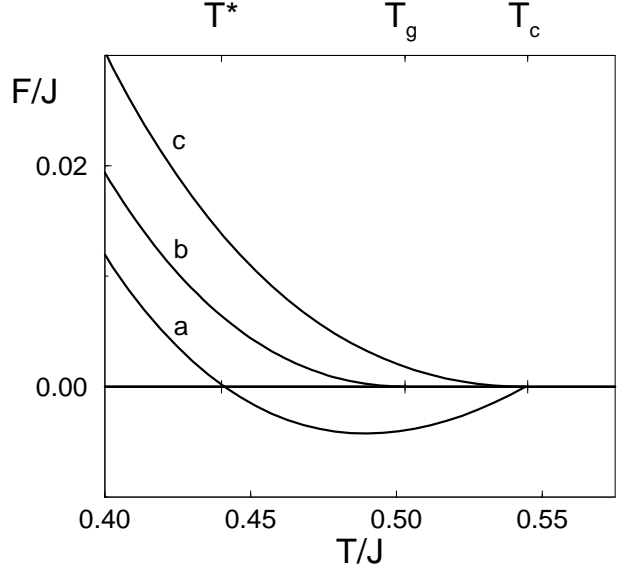,width=9cm}
\caption{
Free energies of a spherical spin glass with random quartet couplings, 
after subtraction of the paramagnetic value.
a) Marginal replica  free energy
b) Static replica free energy
c) ``Experimental'' free energy, obtained by integrating $C/T$
and by analysis of the degeneracy of the TAP states.
}\end{figure}

A simple calculation shows that the `experimental' free energy of CHS and CS,
 and the marginal replica free energy
obtained from eq. (\ref{bFCS})~\cite{maxmin} have the following
connection with the component average free energy ${\overline F}=Nf(q_1)$:
\begin{eqnarray}\label{fdynfeff}
F_{exp}&=&{\overline F}-T{\cal I}_c\\
F_{repl}&=&{\overline F}-T{\cal I}={\overline F}-\frac{T{\cal I}_c }{x(T)}
\label{Frepl}
\end{eqnarray}
Since $x(T_c)=1$ both expressions are at $T_c$ equal to the paramagnetic free 
energy.

In order to trace back the difference between 
(\ref{fdynfeff}) and  (\ref{Frepl}) we have decided to redo the analysis
of the TAP equations. Hereto we consider the generalized partition sum
\begin{equation}\label{Zu=}
Z_u=\sum_a e^{-u\beta F_a(T)}\equiv e^{-\beta F_u}
\end{equation}
 For $u=0$ we thus calculate the
total number ${\cal N}$ of TAP-states, while for $u=1$ we consider their 
partition sum.
The sum over the TAP states can be calculated using standard approaches.
\cite{BM}\cite{Rieger} 
A 1RSB pattern is assumed for
the 6 order parameters. For instance, $q_{\alpha\beta}=(1/N)\sum_i
m_i^\alpha m_i^\beta$ takes the values $q_d$ for $\alpha=\beta$ and
$q_1$ for $\alpha\neq\beta$ both inside a $\tilde x\ast \tilde x$ 
diagonal block
of the 1RSB Parisi matrix, while vanishing outside these blocks.
At fixed breakpoint $\tilde x$ the 12-dimensional saddle point
can be found explicitly.
For the long time limit of the
dynamical approach the marginality condition 
should be taken,~\cite{maxmin} in the form given in eq. (\ref{marg}).
As expected, the above replica expression for $q_1$ is found back as 
solution of $\partial f(q_d)/\partial q_d=0$ at $q_d=q_1$.
The result $q_1=q_d$ asserts that the mutual overlap
between different states in the same cluster is equal to the selfoverlap.
Like in the replica calculation of the ordinary partition sum,
$\tilde x$ can still take any value. In analogy with the marginal 
replica calculation of
eq (2), we expect $\tilde x$ to be fixed by the vanishing of a fluctuation
eigenvalue. We have therefore analyzed that $12\times 12$ longitudinal
fluctuation matrix at marginality. For any value of $\tilde x$ 
it automatically has 3 zero eigenvalues, proving the marginality. 
Another eigenvalue vanishes
for $\tilde x=0$, $\tilde x=1$ (twice) and for $\tilde x=x(T)/u$.  

From this we infer that $\tilde x=x(T)/u$,
and thus ${\cal I}=u{\cal I}_c /x(T)$. In our case $u=1$
it just implies that the calculated complexity is 
the replica value ${\cal I}$, and not ${\cal I}_c$, the one of CS.
This leads to the conclusion that nothing went wrong in the replica
calculation of the dynamical phase transition: 
the replica free energy is  a generating 
function for the mean field equations, and its saddle point value 
is the logarithm of the partition sum.

This conclusion has been supported by a calculation for  the spherical
$p$-spin glass in a transverse field $\Gamma$. ~\cite{ThmNunp} 
In that extension of the model there again occurs a dynamical 
transition from the paramagnet to a 1RSB spin glass state,
 at transition temperature $T_c(\Gamma)$. 
The paramagnet of this model is non-trivial. There is a first order
transition line (that we call {\it pre-freezing line} ) separating
regions with large and small ordering in the $z$-direction.
~\cite{DobrThirum}. 
This line intersects the dynamical
PM-1RSB transition line at a point ($T^\ast,\Gamma^\ast$). 
Beyond this point there occurs a first order PM-SG transition
with a finite latent heat.
We expect the location of the transition line to follow from
matching of free energies. The replica free energy is indeed suited for
that, while the `experimental' free energy does not lead to a meaningful
match.

The free energy is $F_{repl}$ is the physical one, in the sense
that it takes into account the correct value of the complexity.
Nevertheless the increase of complexity, ${\cal I}\sim 1/T$
for low $T$, remains to be explained.

\subsection{Specific heat versus energy fluctuations}

It would be nice to have a measurable quantity that probes
the multitude of states. One object that should be accessible, at least
numerically, is the specific heat. The standard expression 
$C=dU/dT=\sum_a d(p_aU_a)/dT$ is likely to differ from the component
average fluctuations of the internal energy:
${\overline C}=\sum_a C_a=
\sum_a p_a dU_a/dT=\beta^2\sum_ap_a\langle \Delta U_a^2\rangle$.
~\cite{Palmer} The interesting
question is whether their difference is extensive. 
Based on experience in a toy model,~\cite{NvR} we think it
generally is in systems with 1RSB. Since in the present model 
the energy fluctuations are too small at $H=0$,~\cite{fef}
 it can only occur in a field.
From the internal energy 
in a small field we obtain
\begin{eqnarray} 
\frac{1}{N}C(T,H)&=&\frac{1}{2}\beta^2J^2(1+(p-1)q^p-pq^{p-1})\nonumber\\
&-&\beta^2H^2\frac{(p-1)^2(p-2)(1-q)^2}{p(pq+2-p)}\end{eqnarray}
On the other hand, a short calculation shows that $C_a=-Td^2F_a/dT^2$
remains only a function of $q_a$ at the marginal point, which
in the present model takes the field-independent value $q_1=q_d$. 
This implies that ${\overline C}$ is field-independent as well, 
thus satisfying the Parisi-Toulouse hypothesis~\cite{PaT}. Interestingly
enough, we find $C<{\overline C}$, whereas Palmer derives the
opposite at equilibrium. Our reversed ``dynamical'' inequality is a new 
result that is due to the marginality. 

The reversed dynamical inequality occurs due to 
non-equilibrium effects. We conjectured  that it 
generally takes place outside equilibrium, 
for instance in cooling experiments 
above $T_c$ in the three dimensional Edwards-Anderson spin glass.
Some numerical support for this behavior was
found, see ~\cite{Riegernum}

\section{Directed polymer in a correlated random potential}

We  introduce a new, simple model with a static glassy transition.
Consider a directed polymer (or an interface without overhangs)
 $z(x)$ in the section $1\le x \le L$ and
$1\le z \le W$ of the square lattice with unit lattice constant. 
In the Restricted Solid-on-Solid approximation  the interface
can locally be flat ($z(x+1)=z(x)$; no energy cost)
or make a single step ($z(x+1)-z(x)=\pm 1$; energy cost $J$);
larger steps are not allowed.
The polymer is subject to periodic boundary conditions
($z(0)=z(L)$) and we allow all values of $z(0)$.

Further there is a random energy cost
$V(z)$ per element of the polymer at height $z$. Note that this is
a correlated random potential, with energy barriers parallel to the 
$x$-axis.

\subsection{The partiton sum}

The partition sum of this system 
can be expressed in the eigenvalues of the tridiagonal
transfer matrix ${\cal T}$
that has diagonal elements $\exp(-\beta V(z))$ and 
off-diagonal elements $\exp(-\beta J)$
\BEQ
\label{Ztrace}
Z=\tr e^{-\beta\H}=\tr {\cal T}^L=\sum_{w=1}^W 
\left(\Lambda_w\right)^L
\EEQ
For a pure system ($V(z)=0$ for all $z$) at temperature $T=1/\beta$
Fourier analysis tells that for small momentum 
$\Lambda(k)= \Re (1+e^{-\beta J+ik})/(1-e^{-\beta J+ik})$
$\approx \exp[-\beta f_B-Dk^2/(2\pi^2)]$ 
with bulk free energy density $f_B$ 
and diffusion coefficient $ D$ 
that can be simply read off and are temperature dependent.

We shall consider the situation of randomly located potential barriers
parallel to the $x$-axis. 
Hereto we assume binary disorder, so $V(z)=0$ with probability
$p=\exp(-\mu)$ or $V(z)=V_1>0$ with probability $1-p$.
Eq. (\ref{Ztrace}) is dominated by the largest eigenvalues.
It is well known that they 
occur due to Lifshitz-Griffiths singlarities.
These are due to lanes of width $\ell\gg 1$ in which all $V(z)=0$,
 bordered by regions with $V(z)\neq 0$. 
These dominant configurations are the ``components'',
 ``TAP states'' or ``metastable states'' of our previous discussion.
The eigenfunction centered around $z_0$
has inside the lane the approximate form $\cos[ \pi (z-z_0)/\ell]$
while it decays essentially exponentially outside due to the disorder.
These states can thus be labeled by $a=(z_a,\ell_a)$. 
 Since $k\to \pi/\ell$ the free energy of this state  follows as
\BEQ
\beta F_{\ell}\equiv -L\ln\Lambda_\ell \approx \beta f_BL+\frac{DL}{2\ell^2}
\EEQ
The typical number of regions with $\ell$ successive sites with $V=0$ is
$\N_\ell=W(1-p)^2p^\ell$. We now choose $W=\exp(\lambda L^{1/3})$
so the states with width $\ell$ have a  configurational entropy
or  complexity
$\I_{\ell}\equiv \ln\N_\ell \approx \lambda L^{1/3}-\mu\ell$.

\subsection{The TAP-partition sum}
For large $L$ we may restrict the partition sum to these dominant
states.
We thus evaluate, instead of 
eq. (\ref{Ztrace}), the `TAP'
 partition sum
\BEA
Z= \sum_{\ell} \N_\ell e^{-\beta F_\ell} \EEA
Note that it is obtained by simply omitting the contributions
of states with low eigenvalue (high free energy). 
The total free energy 
\BEQ 
\beta F=-\ln Z= \beta f_BL-\lambda L^{1/3}+\mu\ell+\frac{DL}{2\ell^2}
\EEQ
has to be optimized in $\ell$. 
The largest $\ell$ which occurs in the system can be estimated 
by setting $\N_\ell\approx 1 $,  yielding
\BEQ \ell_{max}=\frac{\lambda L^{1/3}}{\mu}
\EEQ      
It is a geometrical length, independent of $T$.
Let us introduce $\tilde D=D\mu^2/\lambda^3$. 
The free energy of this  state reads
\BEQ
\label{Fliq}
\beta F= \beta f_BL+\frac{1}{2}\lambda L^{1/3} \tilde D
\EEQ
At low enough $T$ the optimal length is smaller than $\ell_{max}$,
\BEQ
\ell=\left(\frac{DL}{\mu}\right)^{1/3}=\tilde D^{1/3}\ell_{max}
\EEQ
The free energy of this phase is
\BEQ \label{bFopt}
\beta F =\beta f_BL+\frac{1}{2}\lambda L^{1/3}(3\tilde D^{1/3}-2)
\EEQ
For $\D>1$ ($T>T_g$) the interface is in an essentially
non-degenerate state.
For $\D<1$ it lies in one of the $\N_\ell\gg 1$ relevant states, which
is reminiscent to a glass. So the model has a glassy transition at $\D=1$.

The internal energy of a state of width $\ell$ is
\BEQ
U_\ell=u_BL+\frac{L}{2\ell^2}\,\frac{\partial D}{\partial \beta}
=u_BL+
\left(\frac{\lambda L}{\D^2}\right)^{1/3}\frac{1}{2}
\frac{\partial \D}{\partial \beta}
\EEQ
At $\D=1$ this coincides with the paramagnetic value, simply because
$\ell\to \ell_{max}$. It is easily checked that the free energy (\ref{bFopt}) 
is a thermodynamic potential, and yields the same value for $U$.
At the transition it  branches off quadratically from (\ref{Fliq}).
In the glassy phase the specific heat 
\BEQ
C=\frac{dU}{dT}=c_BL+\frac{L}{2\ell^2}
\partial_T\partial_\beta D+\frac{1}{3}
(\lambda L \tilde D^{-5})^{1/3} (T\partial_T \tilde D)^2 
\EEQ
exceeds the component averaged specific heat 
${\overline C}=Lc_B+(L/2\ell^2)\partial_T\partial_\beta D$.
In contrast to previous model, the specific heat 
it is larger in the glassy phase than in the paramagnet. This is because 
the free energy is lower. 

\subsection{On overlaps and hierarchy of phase space}
  
In a given realization of disorder we define the `overlap'
of two states $a$ and $b$, centered around $z_a$ and $z_b$, respectively,
 as
\BEQ
q_{ab}=\lim_{t\to\infty}\langle \delta_{z(0),z_a}\delta_{z(t),z_b}\rangle
\EEQ
 In the high temperature phase there is 
one non-degenerate state, so $P(q)=\delta(q-q_1)$.
In the glassy phase we expect that $q_{ab}=q_1$ for all 
optimal states $(a,b)$ at temperature $T$ . The reason is  that
at thermodynamic equilibrium the whole phase space
can be traversed, and negligible time is spent in non-optimal states.
If so, then though there are many states, one still has
$P(q)=\delta (q-q_1)$ and there is no replica symmetry breaking. 
This is standard for equilibrium situations without frustration.

This puts forward the picture of replica symmetry breaking
and hierarchy of phase space being a dynamical effect. 
At given timescale only some nearby states can be reached, 
``states within the same cluster''. At larger times
other clusters can be reached, and for times
larger than the ergodic time of a large but finite system,
all states are within reach. Only in the thermodynamic limit
phase space splits up in truely disjoint sets.
To investigate the validity of this picture in detail,
 one should solve the dynamics of the polymer.

\subsection{The polymer model at $\tilde T=1/T$}

The comparison to the $p$-spin model is most direct when we compare
the $p$-spin model at temperature $T$ with the polymer model at temperature
$\tilde T=1/T$. In this interpretation, coming from high $\tilde T$,
 the polymer undergoes a gradual freezing into TAP states.
This truely becomes relevant when the domain size is of order
$\ell\sim L^{1/3}$, where the complexity starts to be smaller than
$\log W=\lambda L^{1/3}$. This gradual freezing
 shows explicitly that the dynamical
transition, as found in the mean field $p$-spin glass, is smeared in finite
dimensions.

For $\tilde T$ going down to $\tilde T_c$,
the polymer gets captured in states with free energy closer and closer
to the lowest free energy state available at that temperature. 

As in 1RSB spin glasses,
the complexity also vanishes to leading order
 for $\tilde T\downarrow \tilde T_c$.
In the low $\tilde T$ phase the complexity is no longer of order $L^{1/3}$.
This is similar to the low $T$ phase of the static $p$-spin model,
where the complexity is non-extensive in the glassy phase.

When considered as function of $\tilde T$, the specific heat makes
a downward jump when cooling the system  from large $\tilde T$ 
below $\tilde T_c$.
The absence of a sharp dynamical transition
and the vanishing of the complexity that occurs in this
polymer model as function of $\tilde T$ are very analoguous to
the expected behavior of realistic glasses.

\subsection*{ Acknowledgments}
The author thanks Yi-Cheng Zhang, H. Rieger and  A. Crisanti for 
stimulating discussion. He is grateful to J.J.M. Franse 
for wise supervision.


\begin{thebibliography}

\bibitem{[1]}{Jackle}{[1]} J. J\"ackle, Phil. Magazine B {\bf 44} (1981) 533
\bibitem{[2]}{Palmer}{[2]} R.G. Palmer, Adv. in Physics {\bf 31} (1982) 669
\bibitem{[3]}{quantummeas}{[3]} This sudden loss of entropy is reminiscent
of the collaps of the wave function in the quantum measurement.
\bibitem{[4]}{KirkpT}{[4]} T.R. Kirkpatrick and D. Thirumalai, 
Phys. Rev. Lett. {\bf 58} (1987) 2091
\bibitem{[5]}{CS}{[5]} 
A. Crisanti and H.J. Sommers, Z. Physik B {\bf 87} (1992) 341
\bibitem{[6]}{CHS}{[6]} A. Crisanti, H. Horner, and H.J. Sommers, 
Z. Phys. B {\bf 92} (1993) 257
\bibitem{[7]}{CK}{[7]} L. F. Cugliandolo and J. Kurchan, Phys. Rev. Lett.
{\bf 71} (1993) 173
\bibitem{[8]}{Gotze}{[8]} E. Leutheusser, Phys. Rev. A {\bf 29} (1984) 2765;
U. Bengtzelius, W. G\"otze, and A. Sj\"olander, J. Phys. C
{\bf 17} (1984) 5915 
 \bibitem{[9]}{Rudi}{[9]} R. Schmitz, J.W. Dufty, and P. De, Phys. Rev.
Lett. {\bf 71} (1993) 2066
\bibitem{[10]}{KPV}{[10]} J. Kurchan, G. Parisi, and M.A. Virasoro, J. Phys. I
(France) {\bf 3} (1993) 1819
\bibitem{[11]}{CSTAP}{[11]} A. Crisanti and H.J. Sommers, J. de Phys. I France
{\bf 5} (1995) 805
\bibitem{[12]}{maxmin}{[12]} Th.M. Nieuwenhuizen, 
 Phys. Rev. Lett. {\bf 74} (1995) 3463
\bibitem{[13]}{Nqsg}{[13]} Th.M. Nieuwenhuizen, 
Phys. Rev. Lett. {\bf 74} (1995) 4289; ibid 4293
\bibitem{[14]}{Rieger}{[14]}H. Rieger, Phys. Rev. B {\bf 46} (1992) 14665
\bibitem{[15]}{KWol}{[15]} T.R. Kirkpatrick and P.G. Wolynes, Phys. Rev.
B {\bf 36} (1987) 8552; D. Thirumalai and T.R. Kirkpatrick,
Phys. Rev. B {\bf 38} (1988) 4881; T.R. Kirkpatrick and D. Thirumalai,
J. Phys. I France {\bf 5} (1995) 777
\bibitem{[16]}{BM}{[16]} A.J. Bray and M.A. Moore, J. Phys. C {\bf 13}
(1980) L469; 
 F. Tanaka and S.F. Edwards, J. Phys. F {\bf 10}
(1980) 2769; 
C. De Dominicis, M. Gabay, T. Garel, and H. Orland,
J. Phys. (Paris) {\bf 41} (1980) 923 
\bibitem{[17]}{ThmNunp}{[17]} Th.M. Nieuwenhuizen, unpublished (1996 )
\bibitem{[18]}{DobrThirum} {[18]} V. Dobrosavljevic and D. Thirumalai,
J. Phys. A {\bf 23} (1990) L767R  
\bibitem{[19]}{NvR}{[19]} 
Th.M. Nieuwenhuizen and M.C.W. van Rossum, Phys. Lett.
A {\bf 160} (1991) 461
\bibitem{[20]}{fef}{[20]} This happens since at $H=0$ 
its (free) energy fluctuations are
not ${\cal O}(\sqrt{N})$ but ${\cal O}(1)$, as can be seen 
by expanding the result for $\ln [Z^n]_{av}$ to order $n^2$.
For $H=0$ there appear no terms of order $n^2N$; 
for $H\neq 0$ they do appear.
\bibitem{[21]}{PaT}{[21]} G. Parisi and G. Toulouse, 
J. de Phys. Lett. {\bf 41} (1980) L361
\bibitem{[22]}{Riegernum}{[22]} 
Such behavior has been observed
for $T>T_g$ in a numerical cooling 
experiment in the 3d Edwards-Anderson model. 
(H. Rieger, private communication, April 1995)
\end{thebibliography}
\end{document}